\documentclass[fleqn,usenatbib,useAMS]{mnras}
\usepackage{amssymb,amsmath}
\usepackage{graphicx,graphics}
\usepackage{xcolor}
\usepackage{multirow}
\usepackage[T1]{fontenc}
\usepackage{ae,aecompl}
\usepackage{newtxtext,newtxmath}
\usepackage{ucs}
\usepackage[utf8x]{inputenc}
\usepackage{epstopdf}

\title[New BLRs radius in Mrk142]
{New approach to broad line region radius in Mrk142 after considering potential short-term 
optical transient quasi-periodic oscillations}

\author[Zhang X. G.]{XueGuang Zhang$^{1}$  
\thanks{Contact e-mail: \href{mailto:xgzhang@njnu.edu.cn}{xgzhang@njnu.edu.cn}}\\
$^{1}$School of Physics and technology, Nanjing Normal University,
          No. 1, Wenyuan Road, Nanjing, 210023, P. R. China}

\pubyear{2020}

\begin{document}
\label{firstpage}
\pagerange{\pageref{firstpage}--\pageref{lastpage}}
\maketitle

\begin{abstract} %%%about 181 words
	Mrk142 has been known as the only outlier in R-L space (correlation between BLRs 
(broad line regions) radii and continuum luminosity) among the low redshift local reverberation 
mapped broad line AGNs (BLAGNs) with moderate accretion rates, due to its BLRs radius smaller 
than R-L expected value. Here, considering probable optical transient quasi-periodic oscillations 
(QPOs), a new approach to assessing the BLRs radius can be considered. Reliable transient QPOs 
in high-energy emissions from black hole vicinity have been reported in several normal AGNs, 
however  there are so-far few short-term low-energy optical transient QPOs in normal BLAGNs (not 
the QPOs reported in blazars nor in AGNs harbouring binary BH systems). Through the photometric 
optical light curves well directly described by sinusoidal functions, we report probable short-term 
optical transient QPOs with periodicities around 14days and 43days in BLAGN Mrk142, indicating 
similar but scaled optical QPOs as those in high-energy bands. Considering the 14days QPOs 
related to reprocessing procedure, new approach to the BLRs radius can be estimated in Mrk142 
through the reverberation mapping technique. The new BLRs radius of Mrk142 well follows the 
R-L relation, demonstrating the R-L relation is fundamental in local normal BLAGNs without 
ultra-high accretion rates.
\end{abstract}


\section{Introduction}

%%%1st
      Transient Quasar-Periodic Oscillations (QPOs) in high-energy emissions arising from general 
relativistic effects (relativistic Frame Dragging method \citep{ref1}, discoseismology method 
\citep{ref2}, etc.) related to central accretion disks have been seen in several Active Galactic 
Nuclei (AGNs) \citep{ref3, ref4, ref5, ref7, ref6}. However, there are so-far rare reports on 
short-term low-energy optical transient QPOs in AGNs with periodicities of about days to tens of 
days, not similar as the reported optical QPOs with periodicities of years to tens of years in 
dozens of blazars \citep{ref8, ref9} related to jet emissions/precessions nor similar as the 
optical QPOs in a few AGNs \citep{ref10} harbouring binary supermassive black hole (BH) systems. 
There are so far few candidates of short-term optical transient QPOs in normal AGNs, such as the 
QPOs in KIC9650712 \citep{ref11} but with low confidence level. Meanwhile, there are several 
reports on probably false optical QPOs in normal AGNs \citep{ref12, ref13, ref14}. Actually, 
either the general relativistic effects related to optical emission regions in accretion disks 
or the reprocessed high-energy emissions \citep{ref15, ref16} to the observed optical emissions 
strongly indicate detectable optical transient QPOs in normal broad line AGNs (BLAGNs). 

\begin{figure*}
\centering\includegraphics[width = 18cm,height=8.5cm]{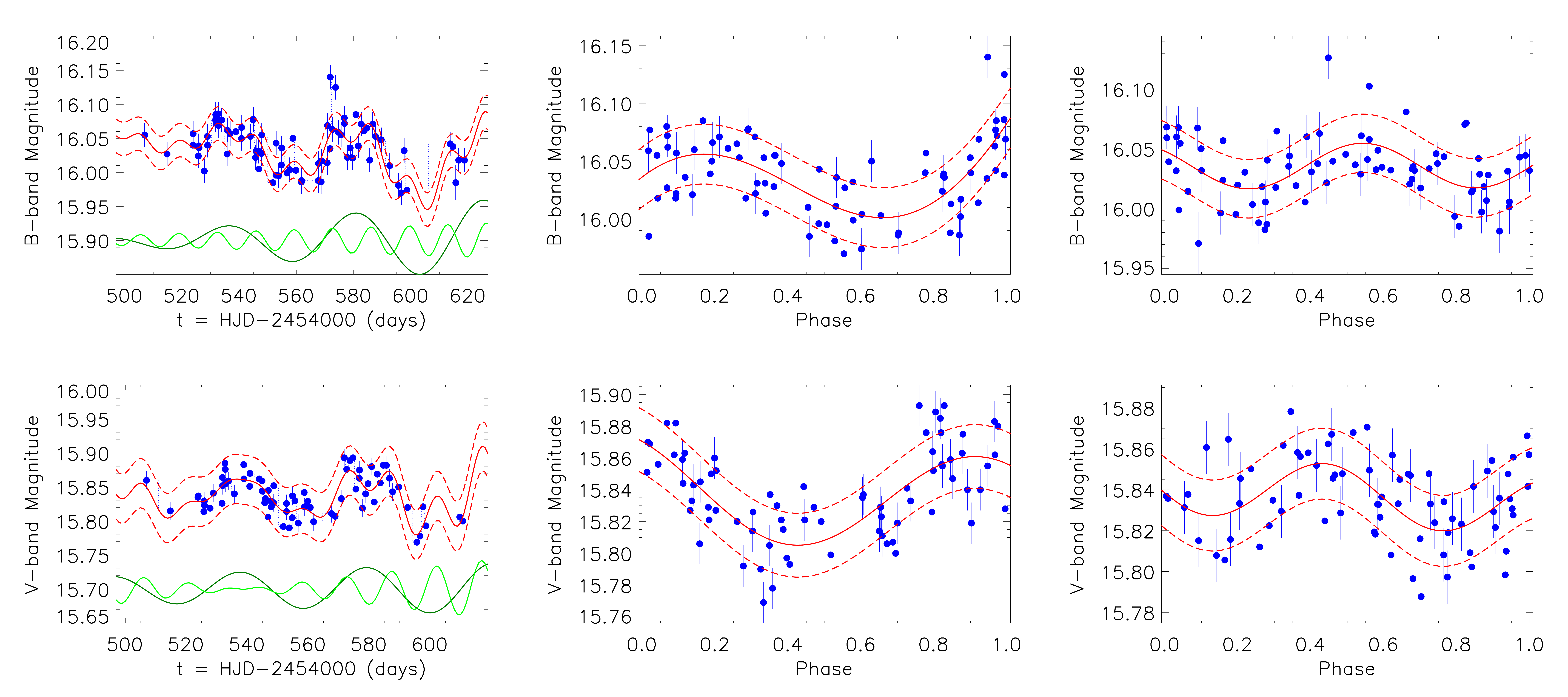}
\caption{Left panels show the observed B-band and V-band light curves (solid blue circles plus 
error bars) and the best-fitting results to each light curve by a linear trend plus two sinusoidal 
functions. Solid lines in dark green and in green show the determined two sinusoidal components 
plus 15.9\ in the top left panel and plus 15.7\ in the bottom left panel. Middle panels show the 
corresponding phase-folded light curves with periodicity of about 41days and the best-fitting 
results to each phase-folded curve by a sinusoidal function plus a linear trend. Right panels show 
the corresponding phase-folded light curves with periodicity of about 16.6days and the best-fitting 
results to each phase-folded curve by a sinusoidal function plus a linear trend, after subtraction 
of the sinusoidal signals with periodicities of 41days from the observed light curves. In each panel, 
solid and dashed red lines show the best-fitting results and the corresponding $1\sigma$ confidence 
bands, respectively. In right panels, due to extra noises from subtractions of the 41days 
quasi-periodic variabilities, the periodicity is about 16.6days in order to find a better sinusoidal 
profile to each phase-folded curve.
}
\label{lmc}
\end{figure*}

%%%2nd
     Timescales of short-term transient QPOs can be scaled with parameters of mass and spin of BH 
and radius of emission regions in different kinds of BH accreting systems \citep{ref17}. From 
accreting systems in galactic X-ray binaries \citep{ref18, ref19, ref20, ref21, ref22} to accreting 
systems in normal BLAGNs, for quasi-periodic emissions from high-energy X-ray band to low-energy 
optical band, expected timescales of short-term optical transient QPOs could be around tens of 
days in normal BLAGNs with central supermassive BHs. To study short-term optical transient QPOs 
in normal BLAGNs could provide further information on structures of large-scale accretion flows 
far-away from central BHs. 

%%%3rd
     Moreover, considering the reverberation mapping method \citep{ref23}, in reverberation 
mapped BLAGNS (RMBLAGNs) \citep{ref24, ref35, ref26, ref27, gd17}, optical continuum (photometric) 
variability properties combining with properties of optical broad emission lines are commonly 
applied to estimate virial BH masses under the Virialization assumption \citep{ref28} to broad 
emission line regions (BLRs). Once short-term optical transient QPOs are detected in optical 
variabilities in normal RMBLAGNs, effects of optical transient QPOs could provide further corrected 
information on structures of BLRs, because optical QPOs will lead to multiple optional 
time lags (with time difference of expected periodicity of the QPOs) between continuum variabilities 
and broad line variabilities as the shown results in the manuscript. 
Here, we report so-far the first unique case with considering effects of probable short-term 
optical transient QPOs on determining BLRs radius in the normal BLAGN Mrk142. The main results 
and discussions are shown in Section 2 on the detected optical QPOs in the well-known RMBLAGN 
Mrk142. Then Section 3 shows our final conclusions.

\begin{figure}
\centering\includegraphics[width = 8cm,height=9cm]{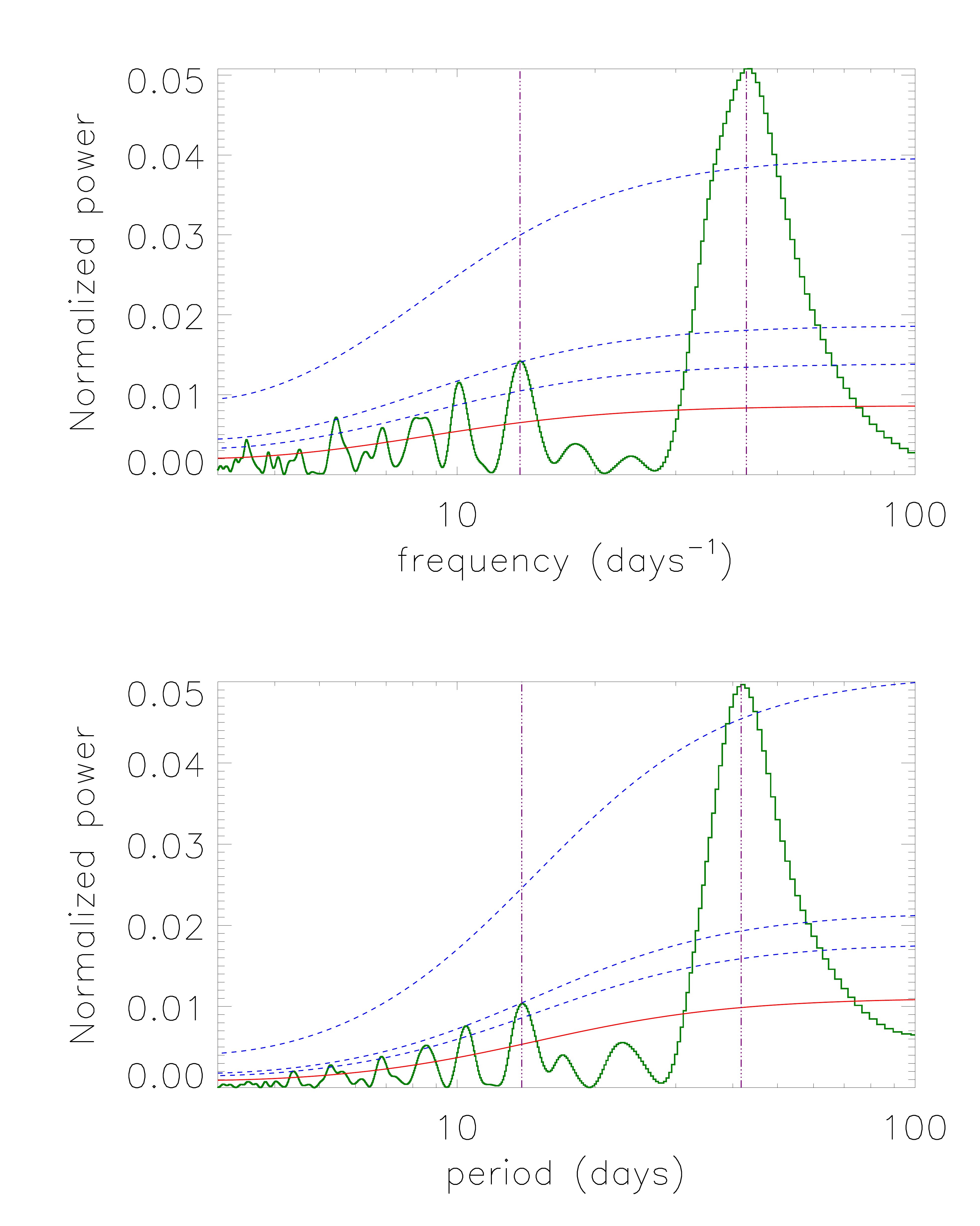}
\caption{The calculated power spectrum by the REDFIT method applied to the observed unevenly
sampled B-band light curve (top panel) and to the observed unevenly sampled V-band light 
curve (bottom panel). In each panel, solid dark-green line represents the results from the 
photometric light curve, solid red line shows the results on the calculated red noise spectrum, 
vertical double-dot-dashed lines in purple mark positions of the two maximum peaks around 14days 
and 41day. In top panel, dashed blue lines from top to bottom represent the $\chi^2$ significance 
levels of 99\%, 88\% and 80\%, respectively. In bottom panel, dashed blue lines from top to 
bottom represent the $\chi^2$ significance levels of 99\%, 85\% and 80\%, respectively. }
\label{redfit}
\end{figure}

\section{Main results and Discussions}

%%%%1st
    Mrk142 is one well-known RMBLAGN in the Lick AGN Monitoring Project (LAMP) \citep{ref25, 
ref29}. Left panels of Fig.~\ref{lmc} show the ground-based high-quality photometric B-band and 
V-band light curves from Feb. to Jun. in 2008 \citep{ref30} in the LAMP, each light curve can 
be well described by one linear trend plus two sinusoidal signals with periodicities around 43days 
and 14days: $T_1\sim13.43\pm0.24{\rm days}$ and $T_2\sim44.78\pm1.25{\rm days}$ for the B-band 
light curve, $T_1\sim14.75\pm0.28{\rm days}$ and $T_2\sim41.56\pm0.82{\rm days}$ for the V-band 
light curve, through the Levenberg-Marquardt least-squares minimization technique. The 
best-fitting results to the observed photometric light curves can be well described by
\begin{equation}
\begin{split}
&LMC(t, B-band)=16.246-3.8\times10^{-4}t+\\
&~~~(0.205-4.2\times10^{-4}t)\times\sin(\frac{2\pi t}{44.78\pm1.25{\rm days}}-1.29)+\\
&~~~(-0.062+1.4\times10^{-4}t)\times\sin(\frac{2\pi t}{13.43\pm0.24{\rm days}}-2.39) \\
&LMC(t, V-band)=15.842-1.1\times10^{-5}t+\\
&~~~(0.061-1.6\times10^{-4}t)\times\sin(\frac{2\pi t}{41.56\pm0.82{\rm days}}-1.11)+\\
&~~~(-0.288+5.3\times10^{-4}t)\times\sin(\frac{2\pi t}{14.75\pm0.28{\rm days}}+2.84)
\end{split}
\end{equation}. 
The best-fitting results lead $\chi^2/DOF$ ($\chi^2$ as the summed squared weighted residuals, 
$DOF$ as the degrees of freedom) to be around 1.8. Actually, the same fitting procedure has been 
applied to the public light curves of the other RMBLAGNs in the projects of LAMP and AGNWATCH, 
Mrk142 is the unique RMBLAGN of which light curves can be well described by sinusoidal functions.

%%2nd
   Meanwhile, the corresponding phase-folded light curves with periodicities of about 41days and 
about 16days (a bit larger than 14days) are shown in the middle and the right panels of Fig.~\ref{lmc}. 
Each phase-folded curve can be well described by a sinusoidal function plus a linear trend. The 
best-fitting results by sinusoidal functions to both the observed and the phase-folded light curves 
indicate apparent QPOs in the optical emissions of Mrk142. Furthermore, amplitude of the sinusoidal 
signal with shorter periodicity included in each observed light curve is quite small, only about 1.5 
times larger than the measured errors of observed light curve, but the well applied statistical 
F-test technique can tell that the sinusoidal signal with shorter periodicity is preferred with 
confidence level higher than 99.6\% ($4\sigma$ confidence levels), rather than by only the sinusoidal 
signal with longer periodicity plus a linear trend applied to describe each observed light curve. 
Similarly, the F-test technique can also determine the confidence level higher than 99.96\% for the 
sinusoidal function plus a linear trend applied to describe each phase-folded light curve, rather 
than by only a linear trend applied.

%%%3rd
    Moreover, the improved REDFIT method \citep{ref32} considering intrinsic AGN variabilities 
dominated by red noises from random walk process \citep{ref33, ref34, ref35} has been well applied 
to check and determine the QPOs included in the observed unevenly sampled B-band and V-band 
light-curves of Mrk142, shown in Fig.~\ref{redfit}. The expected periodicities are around 43days and 
14days through the REDFIT method, similar as the values through the best fitted results directly 
by sinusoidal functions shown in Fig.~\ref{lmc}. Meanwhile, as the shown results in 
Fig.~\ref{redfit}, there seems to be a possible periodicity of 10days, at equal confidence levels 
to the 14days. However, if 10days rather than the 14days applied in the shorter QPO signals, the 
calculated $\chi^2/DOF$ should be larger than 2, larger than the $\chi^2/DOF\sim1.8$ with 14days as 
the shorter periodicity. Furthermore, if 10days was considered as the shorter periodicity, there 
should be an apparent drop of the confidence levels from 99.96\% to 80\% for the sinusoidal function 
plus a linear trend applied to describe the phase-folded light curves with the shorter periodicity 
of 10days applied (such as the ones shown in the right panels of Fig~\ref{lmc}), rather than by 
only a linear trend applied. Therefore, the periodicity of 14days is preferred in the manuscript. 
Based on the results through the REDFIT method, the $\chi^2$ significance levels for the periodicities 
around 14days and 43days are higher than 85\% and 99\%, indicating the quasi-periodic variabilities 
are reliable to some extent. Meanwhile, based on the power spectra in Fig.~\ref{redfit}, there are 
much different qualities $Q_p~=~p/\delta~p$ ($p$ and $\delta~p$ as the periodicity and the 
corresponding full width at half maximum) of the two periodicities, $Q_p~\sim~2.1$ and $Q_p~\sim~6.1$ 
for the periodicity of around 43days and of around 14days, respectively, indicating the two 
periodicities are related to emissions from different emission processes and/or to emissions 
from different emission regions.

%%%4th
  Particularly based on the results from the best-fitting results directly by the sinusoidal signals 
shown in Fig.~\ref{lmc}, the optical transient QPOs can be detected, especially the quasi-periodic 
variability component with periodicity around 14days due to eight complete cycles covered in the light 
curves, after considerations of the time durations and time cadences of the light curves from the LAMP. 
Why are there two different kinds of short-term optical quasi-periodic signals in the normal BLAGN 
Mrk142? The answer is simple but fascinating. After considering the well applied reprocessing procedure 
\citep{ref15, ref16} in accreting system around the central supermassive BH in Mrk142, the QPOs with 
longer periodicity come from local optical continuum emissions with the emission regions in outer part 
of the central accretion disk, however the QPOs with shorter periodicity come from the reprocessed 
X-ray band emissions from the inner accretion disk and/or from the central coronal regions. Therefore, 
it is reasonable to find short-term optical weeks-timescale quasi-periodic variabilities in Mrk142.

\begin{figure}
\centering\includegraphics[width = 8cm,height=5.5cm]{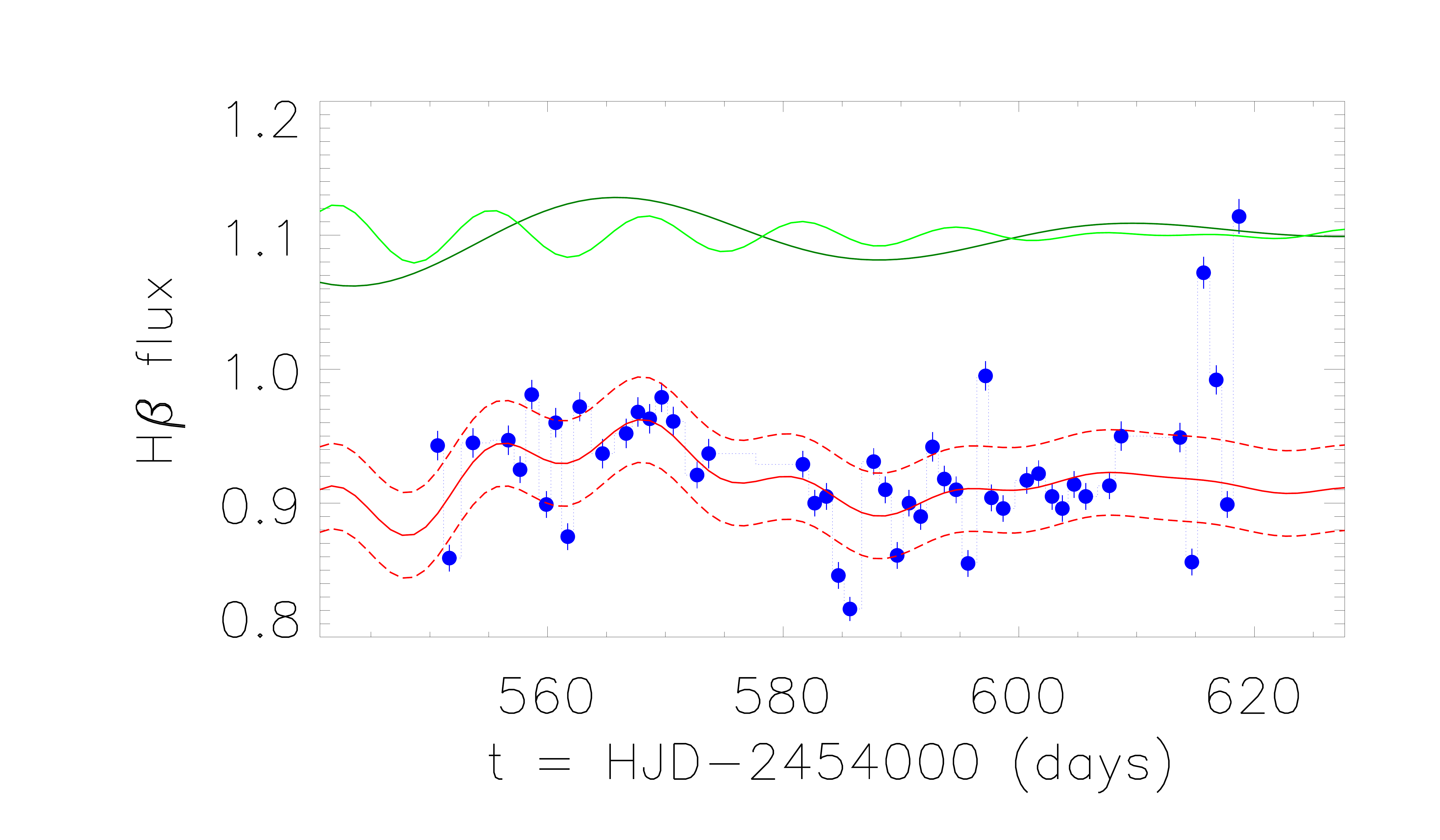}
\caption{The light curve (solid blue circles plus error bars) of broad H$\beta$ collected from 
the LAMP, and the corresponding best-fitting results. Line styles and symbols have the same meanings 
as those applied in the left panels of Fig.~\ref{lmc}. Solid lines in dark green and in green show 
the determined two sinusoidal components plus 1.1.}
\label{hb}
\end{figure}

\begin{figure*}
\centering\includegraphics[width = 18cm,height=9cm]{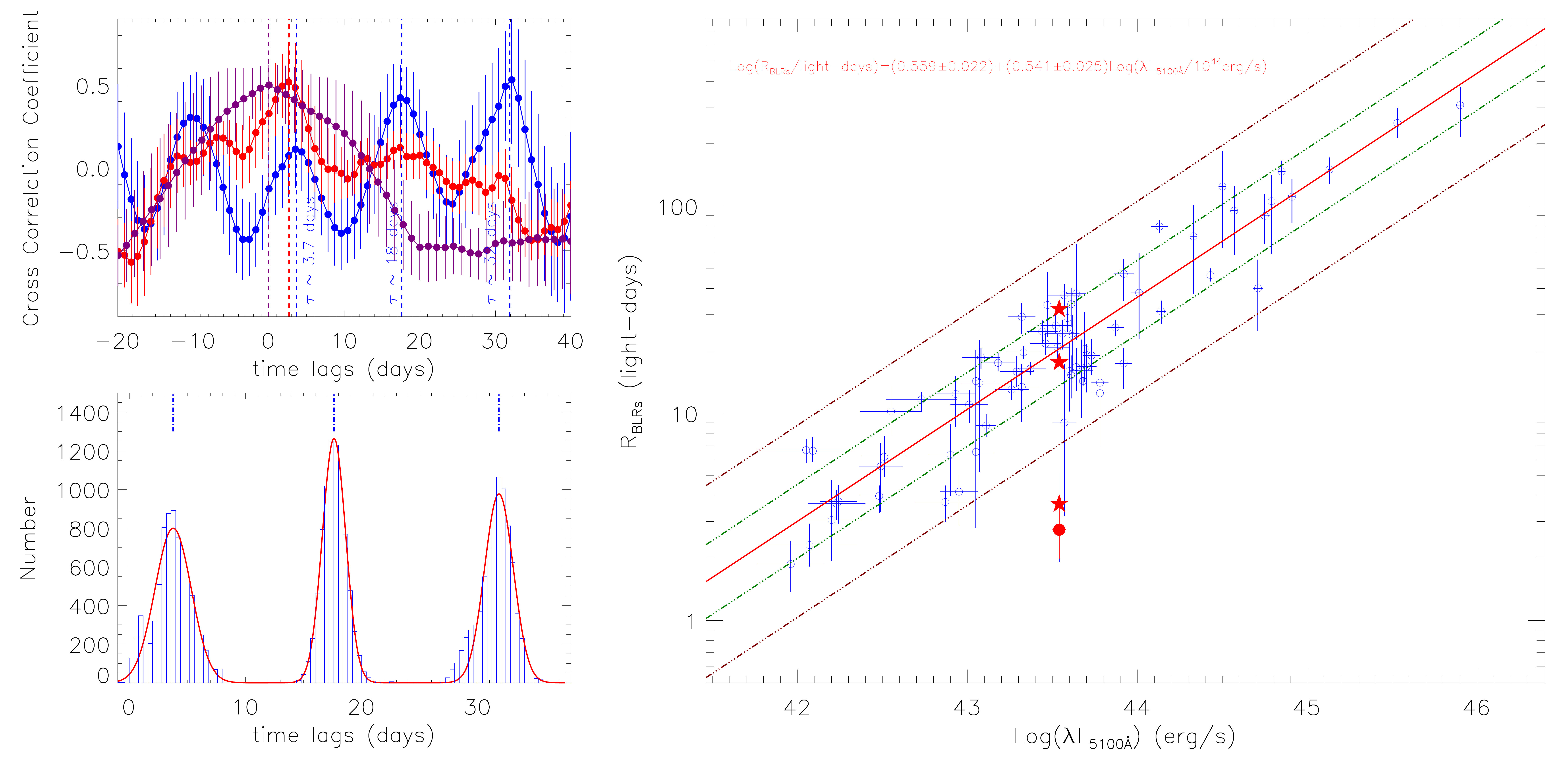}
\caption{Top left panel shows the cross correlation results and the corresponding uncertainties 
determined through the Bootstrap method. Symbols and lines in red, blue and purple show the cross 
correlation results of $CCF_{\rm ori}$, $CCF_{\rm QPO14}$ and $CCF_{\rm QPO43}$, respectively. 
Vertical thick dashed lines in red and in blue mark positions of the peak in the $CCF_{\rm ori}$ 
and the three positive peaks in the $CCF_{\rm QPO14}$, respectively. Vertical thick dashed purple 
line marks the position of zero time lag. Bottom left panel shows peak position distributions of 
the three positive peaks in $CCF_{\rm QPO14}$ around 3.7days, 18days and 32days by the Bootstrap 
method. Solid red lines show the Gaussian fitted results to the distributions. Vertical dot-dashed 
blue lines show the mean value positions of the three positive peaks. Right panel shows the R-L 
relation for the 71 data points from the 41 well-known low redshift local RMBLAGNs in the sample 
of \citet{ref25}. Solid red circles shows the position of Mrk142 with the BLRs radius of 
2.7light-days previously reported in the LAMP, an apparent outlier in the R-L space. The three 
solid red five-point-stars from bottom to top show the new determined values of $3.7\pm1.5$days, 
$17.6\pm1.2$days and $31.6\pm1.4$days in Mrk142. Solid lines in red, dark green and dark red show 
the best-fitting results and the corresponding 68\% and 99\% confidence bands, respectively. 
The formula of the best-fitting results is shown in the top left corner.}
\label{ccf}
\end{figure*}

%%%5th
   Considering the reprocessing procedure, more meaningful results on radius of broad line emission 
regions (BLRs) can be expected in Mrk142 through the known reverberation mapping technique \citep{ref23, 
ref28}, because broad emission line variabilities more sensitively depend on high-energy photoionization 
photons, i.e., photons in the quasi-periodic emissions with shorter periodicities. In Mrk142, the BLRs 
radius has been reported as 2.7light-days \citep{ref25} or 6light-days \citep{ref27} in different epochs 
by time lags between broad line emissions and continuum emissions. However, considering the commonly 
accepted R-L relation (the correlation between BLRs radius and continuum luminosity) \citep{ref25} with 
the optical continuum luminosity of Mrk142 in LAMP, its expected BLRs radius around 20light-days is about 
7 times larger than the reported 2.7light-days in the LAMP, leading Mrk142 to be the only outstanding 
outlier in the R-L space among the known RMBLAGNs with moderate accretion rates in the sample of 
\citet{ref25}. Moreover, besides the reported RMBLAGNs in \citet{ref25}, \citet{dz18} have reported 
the SEAMBH RMBLAGNs with high accretion rates, and \citet{gd17} have reported the SDSS RMBLAGNs with 
redshift around 0.5. And a larger part of SEAMBH RMBLAGNs and SDSS RMBLAGNs have apparently smaller BLRs 
radii than the R-L relation expected values. However, as discussed in \citet{ft20} and in \citet{dz18}, 
the smaller BLRs radii in SEAMBH RMBLAGNs and SDSS RMBLAGNs could be due to super-Eddington accretion 
process and/or tightly related to quite different UV/optical spectral energy distributions. Here, among 
the low redshift local RMBLAGNs with moderate accretion rates in the \citet{ref25}, Mrk142 was the 
definite outlier in the R-L space. Therefore, it is interesting and thoughtful to consider whether are 
there intrinsic reasons applied to well explain the reported BLRs radius of Mrk142 smaller than the R-L 
expected value, besides the well-considered effects of super-Eddington accretion process \citep{ref39, 
dz18}. Here, the quasi-periodic high-energy emissions have been determined with periodicity around 14days 
in Mrk142, considering the reprocessing procedure. Therefore, it is interesting to consider and check 
effects of the QPOs in Mrk142 on the final determined time lag of the broad line emissions relative 
to the continuum emissions.

%%%%6th

    Before proceeding further, it is interesting to quickly check whether are there QPOs included in 
the broad H$\beta$ variabilities in Mrk142. Fig.~\ref{hb} shows the collected light curve of broad H$\beta$ 
in Mrk142 from the LAMP \citep{ref26}. And then, the same linear trend plus two sinusoidal functions have 
been applied to describe the light curve. Similar two sinusoidal components can be expected with 
periodicities of about $45\pm4$days and $13.2\pm0.5$days, however, the calculated $\chi^2/DOF$ value is 
about 18. The determined sinusoidal components but large $\chi^2/DOF$ indicate that there are signs to 
support the expected QPOs in Mrk142, but the results from the broad H$\beta$ variabilities have lower 
confidence levels. Therefore, the cross correlation results on the broad H$\beta$ variabilities are 
mainly considered as follows.

%%%%7th
   The direct cross correlations are shown in top left panel of Fig.~\ref{ccf} not only between the 
observed photometric B-band emissions and the observed broad H$\beta$ emissions (the results called as 
$CCF_{\rm ori}$), but also between the observed broad H$\beta$ emissions and the continuum emissions 
described only by the quasi-periodic emissions with the shorter periodicity of about 14days (the component 
shown as solid green line in the top-left panel of Fig.~\ref{lmc}) (the results called as $CCF_{\rm QPO14}$) 
and between the observed broad H$\beta$ emissions and the continuum emissions described only by the 
quasi-periodic emissions with the longer periodicity of about 43days (the component shown as solid 
dark-green line in the top-left panel of Fig.~\ref{lmc}) (the results called as $CCF_{\rm QPO43}$). 
Based on the cross correlation results in the top-left panel of Fig.~\ref{ccf}, three points can 
be well confirmed. First, there are no apparent time lags in $CCF_{\rm QPO43}$. Zero time lag in 
$CCF_{\rm QPO43}$ can be well expected, because the QPOs with longer periodicity come from optical 
emission regions, but the broad H$\beta$ emissions are tightly related to high ionization photons (the 
QPO emissions with shorter periodicity). Therefore, to check the time lags in the $CCF_{\rm QPO14}$ 
could be preferred. Second, the time lag can be well confirmed to be around 2.7days with maximum cross 
correlation coefficient about 0.5\ in $CCF_{\rm ori}$, the same as the results in the literature 
\citep{ref26}, indicating reliable applications of our procedure to determine time lags. Meanwhile, 
in $CCF_{\rm QPO14}$, the first peak with positive time lag is about 3.7days, much near to the 2.7days 
determined in the $CCF_{\rm ori}$. Third, due to properties of the light curve described by the QPOs 
with shorter periodicity, multiple peak positions could be expected in the corresponding $CCF_{\rm QPO14}$. 
The expected multiple peaks by properties of QPOs, here 4 peaks, can be found and well determined 
around $(3.7~\pm~k~\times~14)$days ($k=-1,~0,~1,~2$) in the $CCF_{QPO14}$.

%%%8th
   Distributions of the three positive time lags in $CCF_{\rm QPO14}$ are shown in the bottom left 
panel of Fig.~\ref{ccf} through the Bootstrap method, leading the determined time lags to be 
$3.7\pm1.5$days, $17.6\pm1.2$days and $31.6\pm1.4$days. Based on properties of the QPOs with shorter 
periodicity coming from the reprocessed high energy emissions, the new preferred BLRs radius of Mrk142 
could be $17.6\pm1.2$days or $31.6\pm1.4$days, now coincident with the expected value by the continuum 
luminosity through the R-L relation, considering the best-fitting results and the corresponding 
confidence bands shown in the right panel of Fig.~\ref{ccf} by the Least Trimmed Squares robust 
technique \citep{ref40}. The new approach to the BLRs radius applied to estimate the central virial 
BH mass of Mrk142 leads to the intrinsic accretion rate at least six times smaller than the previous 
determined value, clearly indicating the BLAGN Mrk142 is definitely not a BLAGN with super-Eddington 
accretion rate, and there are few effects of ultra-high accretion rates on the BLRs radius expected 
through the continuum luminosity. The results ensure no one outlier in the R-L space among the 
reported local RMBLAGNs with moderate accretion rates in the sample of \citet{ref25}, therefore, 
the R-L relation can be well accepted as a fundamental relation in normal BLAGNs.

	Before the end of the section, there is one main point we should note. With more and more 
RMBLAGNs occupying broader parameter space, a growing number of RMBLAGNs will have their BLRs radii 
that differ significantly from the commonly accepted R-L relation expected values. As more recent 
discussed results in \citet{ft20}, further understanding on intrinsic properties of ionizing continuum 
emissions could well explain the R-L offsets. However, as the shown special results with applications 
of detected QPOs well leading Mrk142 not to be an outlier in the R-L space any more, it is worthwhile 
searching optical QPOs candidates in several rare RMBLAGNs with smaller BLRs radii. Certainly, potential 
QPOs can not be commonly applied to explain the global R-L offsets in RMBLAGNs, nevertheless, the 
shown results in the manuscript will lead a new way to discuss optical QPOs in normal AGNs.

\section{Conclusions}

      Based on the months-long high-quality photometric B-band and V-band light curves of Mrk142, 
short-term optical transient QPOs in Mrk142 could be expected with significance levels higher than 
85\%, with periodicities of about 14days, indicating relativistic effects could be detected in 
low-energy optical emissions through the reprocessed emissions. Moreover, considering effects of 
the probable short-term optical transient QPOs, improved properties of BLRs can be estimated  
through the reverberation mapping technique widely applied in normal BLAGNs, leading to the new 
BLRs radius in Mrk142 which is previously known as the only outlier among the low redshift local 
RMBLAGNs with moderate accretion rates. The new BLRs radius of Mrk142 well follows the R-L relation, 
indicating the R-L relation is fundamental enough in normal local BLAGNs without ultra-high 
accretion rates.

\section*{Acknowledgements}
Zhang gratefully acknowledge the anonymous referee for giving us constructive comments 
and suggestions greatly improving our paper. Zhang gratefully acknowledges the kind support 
of Starting Research Fund of Nanjing Normal University, and the kind grant support from 
NSFC-11973029.

\section*{Data Availability}
The data underlying this article will be shared on reasonable request to the corresponding author 
(\href{mailto:xgzhang@njnu.edu.cn}{xgzhang@njnu.edu.cn}).

\label{lastpage}
\end{document}